%Paper: gr-qc/9301007
%From: yoshiaki ohkuwa <ohkuwa@cosmic.physics.ucsb.edu>
%Date: Sat, 9 Jan 93 12:32:38 -0800

\input harvmac

\def\ajou#1&#2(#3){\ \sl#1\bf#2\rm(19#3)}
\def\Title#1#2{\rightline{#1}\ifx\answ\bigans\nopagenumbers\pageno0\vskip1in
\else\pageno1\vskip.8in\fi \centerline{\titlefont #2}\vskip .5in}

scaled\magstep3
 
scaled\magstep3
\font\ticp=cmcsc10

						%	references
\def\Ref#1{Ref.\thinspace#1}                    %       ditto
     % For citation of equation numbers
     	        %       ditto
                    %       ditto
                    %       ditto
                  %       ditto
                  %       ditto
\def\frac#1#2{{#1 \over #2}}
%
%References
%
\lref\GH{M.~Gell-Mann and J.B.~Hartle in {\sl Complexity, Entropy,
and the Physics of Information, SFI Studies in the Sciences of
Complexity}, Vol.
VIII, ed. by W.~Zurek (Addison Wesley, Reading, 1990)
or in {\sl Proceedings
of
the 3rd
International Symposium on the Foundations of Quantum Mechanics in the
Light of
New Technology}, ed.~by S.~Kobayashi, H.~Ezawa, Y.~Murayama, and
S.~Nomura
(Physical Society of Japan, Tokyo, 1990).}
\lref\Gri{R.B.~Griffiths, \ajou J.~Stat.~Phys. &36 (84) 219.}
\lref\Omn{R.~Omn\`es, \ajou J.~Stat.~Phys. &53 (88) 893;
\ajou  ibid. &53 (88) 933;
\ajou ibid. &53 (88) 957; \ajou Ann.~Phys.~(N.Y) &201 (90) 354.}
\lref\YT{N.~Yamada and S.~Takagi, \ajou Prog.~Theor.~Phys. &85
 (91) 985.}
\lref\Eve{H.~Everett, \ajou Rev.~Mod.~Phys. &29 (57) 454; Ref. 6, P.3.}
\lref\DG{B.S.~DeWitt and N.~Graham, {\sl The Many-Worlds
Interpretation of Quantum Mechanics}
(Princeton Univ.~Press, Princeton, 1973).}
\lref\Whe{J.A.~Wheeler, \ajou Rev.~Mod.~Phys & 29 (57) 463.}
\lref\Gra{N.~Graham, Ref. 6, P.229.}
\lref\Har{J.B.~Hartle, \ajou Am.~J.~Phys. &36 (68) 704.}
\lref\DeWa{B.S.~DeWitt, \ajou Physics Today &23 (70) no.~9.}
\lref\DeWb{B.S.~DeWitt, Ref. 6, P.167.}
\lref\FGG{E.~Farhi, J.~Goldstone, and S.~Gutmann, \ajou
Ann.~Phys.~(N.Y.) &192 (89) 368.}

\line{\hfill UCSBTH-92-40}
\line{\hfill MMC-M-2}
\line{\hfill October 1992}
\vskip .26 in
\centerline{\bf Decoherence Functional and Probability Interpretation}
\vskip .50 in
\centerline{\ticp Yoshiaki Ohkuwa}
%\footnote{$^*$}{e-mail:
%ohkuwa@cosmic.physics.ucsb.edu}}
\vskip .26 in
\centerline{\sl Department of Mathematics, Miyazaki Medical College}
\centerline{\sl Kiyotake, Miyazaki 889-16 JAPAN}
\vskip .13 in
\centerline{\sl and}
\vskip .13 in
\centerline{\sl Department of Physics, University of California}
\centerline{\sl Santa Barbara, CA 93106-9530 USA}
\vskip .50 in

{
\midinsert\narrower\narrower
\centerline{\bf ABSTRACT}

We confirm that the diagonal elements of the Gell-Mann and Hartle's
decoherence functional are equal to the relative frequencies of the
results of many identical experiments, when a set of alternative
histories decoheres.  We consider both cases of the pure and mixed
initial states.

\endinsert
}
\vfill\eject
\noindent{\bf \S~1. Introduction}

If quantum mechanics is the fundamental theory of physics, the entire
universe should also be described quantum mechanically.  Recently
Gell-Mann and Hartle \refs{\GH} generalized the quantum theory using the
concept of coarse graining and decoherence.  Similar frameworks were
constructed primarily by Griffiths
\refs{\Gri} and Omn\` es \refs{\Omn}, and
Yamada and Takagi \refs{\YT} constructed independently the similar
framework.  They showed that, when a set of alternative histories
decoheres, the diagonal elements of the decoherence functional satisfy
the mathematical properties of probabilities, so they regarded these as
physical probabilities.

On the other hand, Everett \refs{\Eve} and others \refs{\DG \Whe \Gra
\Har \DeWa \DeWb - \FGG}
discussed that probability interpretation of the quantum theory can be
explained by the fundamental framework of the theory itself, though a
``measure'' in Hilbert space is introduced.  They considered the relative
frequencies of the results of many ($N$) identical experiments.  They showed
that the absolute squares which are identified with probabilities in
ordinary quantum mechanics are equal to the relative frequencies, when
$N\to \infty$.

In this paper we confirm that the diagonal elements of the Gell-Mann and
Hartle's decoherence functional are equal to the relative frequencies,
when a set of alternative histories decoheres and $N\to \infty$.  We
consider the pure state case in \S 2 and the mixed state case in \S
3.
\vskip .13 in
\noindent{\bf \S~2. Pure State Case}

In order to consider probability interpretation we need an ensemble of
identical systems.  In this section an individual system  is a pure
state $|\psi\rangle$, and $|\psi\rangle$ is normalized to unity,
$\langle\psi|\psi\rangle = 1$.  Suppose we have $N$ identical systems,
so that the total system is described by the state vector,

\eqn\one{
|\Psi\rangle = |\psi\rangle^N = |\psi\rangle \cdots |\psi\rangle
\qquad (N\ {\rm terms})\ .
}

Following Gell-Mann and Hartle \refs{\GH} let us consider histories,

\eqn\two{
C_\alpha = P^n_{\alpha_n} (t_n) \cdots P^1_{\alpha_1} (t_1) \ .
}
Here $P^k_{\alpha_k}(t_k)$ $(k=1, \cdots, n)$ are projection operators
and satisfy

\eqn\three
{\sum_\alpha P^k_\alpha (t) = 1\quad , \quad P^k_\alpha (t) P^k_\beta
(t) = \delta_{\alpha\beta} P^k_\alpha (t) \ .
}
In $P^k_\alpha(t)$, $k$ labels the set, $\alpha$ the particular
alternative, and $t$ its time.

The decoherence functional can be written as

\eqn\four{
\eqalign{D\left(C_{\alpha^\prime}, C_\alpha\right) & = Tr
\left[C_{\alpha^\prime} \rho\ C^\dagger_\alpha\right]\cr
&= Tr \left[P^n_{\alpha^\prime_n} (t_n) \cdots
P^1_{\alpha^\prime_1}(t_1)\, \rho\, P^1_{\alpha_1}(t_1) \cdots
P^n_{\alpha_n} (t_n)\right]\ ,\cr}
}
where $\rho$ is the initial density matrix,

\eqn\five{
\rho = |\psi\rangle\langle \psi | \ .
}
A set of histories, $C_\alpha, C_{\alpha^\prime}$ is said to decohere,
when

\eqn\six{
D\left(C_{\alpha^\prime}, C_\alpha\right) = 0 \qquad \left( {\rm for
\ any}\ \alpha^\prime_k \not= \alpha_k\right) \ .
}

In the following discussion we assume that the set of histories
decoheres. From Eqs.~\four -- \six~ we obtain

\eqn\seven{
\eqalign{
D\left(C_{\alpha^\prime}, C_\alpha\right) &= \sum_{\psi^\prime}
\left\langle\psi^\prime \left| C_{\alpha^\prime} \right| \psi
\right\rangle \left\langle\psi\left| C^\dagger_\alpha \right|
\psi^\prime\right\rangle\cr
&= \left\langle\psi\left| C^\dagger_\alpha C_{\alpha^\prime} \right|
\psi \right\rangle\cr
&= \delta_{\alpha\alpha^\prime} \left\langle\psi \left| C^\dagger_\alpha
C_\alpha \right| \psi \right\rangle\ ,\cr
{\rm with} \ \delta_{\alpha\alpha^\prime} &= \prod^n_{k=1}
\delta_{\alpha_k \alpha^\prime_k}\ .\cr}
}

Denote the diagonal element of decoherence functional as $P[C_\alpha]$,
that is

\eqn\eight{
P\left[C_\alpha\right] = D\left(C_\alpha, C_\alpha\right)\ .
}
Starting from  $\sum_\beta\langle \psi | C^\dagger_\beta C_\alpha | \psi
\rangle$ or $\sum_\beta \langle\psi | C^\dagger_\alpha C_\beta | \psi
\rangle$, we can see

\eqn\nine{
P\left[C_\alpha\right] = \left\langle\psi \left | C^\dagger_\alpha
C_\alpha \right| \psi \right\rangle = \left\langle\psi \left | C_\alpha
\right | \psi \right\rangle = \left\langle\psi \left | C^\dagger_\alpha
\right| \psi \right\rangle\ .
}
It is easy to see that $P[C_\alpha]$ satisfy the axiom of mathematical
probability:

\eqn\ten{
P\left[C_\alpha\right]\geq 0, \quad \sum_\alpha
P\left[C_\alpha\right]=1, \quad P\left[C_\alpha + C_\beta\right] = P
\left[C_\alpha\right] + P\left[C_\beta\right]\ (\alpha\not=\beta)
\ .
}
We will show that these $P[C_\alpha]$ are equal to the relative
frequencies of the results of many $(N)$ identical experiments.

Consider that histories $C_{\alpha^I} = P^n_{\alpha^I_n} (t_n)
\cdots P^1_{\alpha^I_1} (t_1) \quad (I=1,\cdots, N)$ are those
of $N$ identical systems, but they may not be same histories.
Here $P^k_{\alpha^I_k} (t_k)\quad
(k=1,\cdots, n)$ act on $I$th factor of Eq.~\one, upper indices of
$\alpha$ distinguish the individual system, and lower indices
distinguish the time slice.

Let us define the relative frequency\footnote{$^*$}{In this case it is
also possible to define an operator which corresponds to the relative
frequency: $\hat F_\alpha = \sum_{\alpha^1\cdots\alpha^N}
|\alpha^1\rangle \cdots | \alpha^N \rangle f_\alpha (\alpha^1, \cdots,
\alpha^N) \langle \alpha^N | \cdots \langle
\alpha^1|$ (See \Ref{\Gra}, {\Har}, {\FGG}).} of $\alpha$ in the sequence
$\alpha^1,\cdots, \alpha^N$ by

\eqn\eleven{
f_\alpha \left(\alpha^1, \cdots, \alpha^N\right) = \frac{1}{N}
\sum^N_{I=1} \delta_{\alpha\alpha^I} \ .
}
And let us define

\eqn\twelve{\delta \left(\alpha^1, \cdots, \alpha^N\right) = \sum_\alpha
\left[f_\alpha \left(\alpha^1, \cdots, \alpha^N\right) -
P\left[C_\alpha\right]\right]^2\ ,
}
which measures the degree to which the sequence $\alpha^1, \cdots,
\alpha^N$ deviates from a random sequence with weights $P[C_\alpha]$.

We write

\eqn\thirteen{
|\alpha\rangle = \frac{C_\alpha |\psi\rangle}{\sqrt{\langle\psi|
C_\alpha |\psi\rangle}}\ ,
}
and we obtain from Eqs.~\two, \three, \seven, \nine

\eqn\fourteen{
\eqalign{
\langle\alpha | \beta\rangle &= \delta_{\alpha\beta}\ ,\cr
|\psi\rangle & = \sum_\alpha\langle \alpha |\psi \rangle | \alpha\rangle
\ , \cr
|\langle\alpha|\psi \rangle |^2 &= P\left[C_\alpha\right]\ .\cr}
}
With these ortho-normal vectors
$|\alpha\rangle$
we can expand the total
wave function as

\eqn\fifteen{
|\Psi\rangle = |\psi \rangle^N = \sum_{\alpha^1 \cdots \alpha^N} \langle
\alpha^1 | \psi \rangle \cdots \left\langle \alpha^N\left| \psi
\bigr\rangle\bigr|\alpha^1 \bigr\rangle \cdots \right| \alpha^N
\right\rangle\ .
}
Let $\epsilon$ be an arbitrarily small positive number and let us define

\eqn\sixteen{
\eqalign{
\big|\Psi^\epsilon_N\bigr\rangle &=
\sum_{\scriptstyle\alpha^1\cdots\alpha^N\atop
\scriptstyle\delta(\alpha^1\cdots
\alpha^N)<\epsilon} \left\langle\alpha^1 |\psi\right\rangle
\cdots \left\langle\alpha^N |\psi\right\rangle\,|\alpha^1\rangle
\cdots |\alpha^N\rangle\ ,\cr
\left| \chi^\epsilon_N \right\rangle &=
\sum_{\scriptstyle\alpha^1\cdots\alpha^N \atop
\scriptstyle\delta(\alpha^1\cdots\alpha^N)\ge\epsilon}
\left\langle\alpha^1 |\psi\right\rangle
\cdots \left\langle\alpha^N |\psi\right\rangle\,|\alpha^1\rangle
\cdots |\alpha^N\rangle\ .\cr}
}
Then from Eqs.~\ten, \eleven, \twelve, \fourteen, \sixteen \ and Eq.~
(4.13) of \Ref \refs{\DeWb}, we can prove that

\eqn\seventeen{
\eqalign{
\left\langle \chi^\epsilon_N | \chi^\epsilon_N \right\rangle &=
\sum_{\scriptstyle\alpha^1\cdots\alpha^N \atop \scriptstyle
\delta (\alpha^1 \cdots \alpha^N) \ge \epsilon}
\left|\left\langle \alpha^1 \big| \psi \right\rangle \right|^2 \cdots
\left|\left\langle \alpha^N \big| \psi \right\rangle \right|^2 \cr
&\leq \frac{1}{\epsilon} \sum_{\alpha^1\cdots \alpha^N} \delta
\left(\alpha^1, \cdots, \alpha^N\right) \left|\left\langle \alpha^1
\big| \psi \right\rangle \right|^2 \cdots \left|\left\langle \alpha^N
\big| \psi \right\rangle\right|^2\cr
&= \frac{1}{\epsilon} \sum_{\alpha\alpha^1\cdots \alpha^N}
\left[f_\alpha \left(\alpha^1, \cdots, \alpha^N\right) - |\langle\alpha| \psi
\rangle |^2 \right]^2
\left|\left\langle \alpha^1 \big| \psi \right\rangle \right|^2 \cdots
\left|\left\langle \alpha^N \big| \psi \right\rangle \right|^2 \cr
&= \frac{1}{\epsilon} \sum_\alpha \frac{1}{N}\ |\langle \alpha | \psi
\rangle |^2 \left(1-|\langle\alpha | \psi \rangle |^2 \right)\cr
&\leq \frac{1}{N\epsilon}\ .\cr}
}
No matter how small we choose $\epsilon$, we can always find an $N$ big
enough so that the norm of $|\chi^\epsilon_N\rangle$ becomes smaller than
any positive number.  This means that

\eqn\eighteen{
\lim_{N\to\infty} |\Psi^\epsilon_N \rangle = |\Psi\rangle\ .
}
Therefore we have shown that $P[C_\alpha]$ are equal to the relative
frequencies.

\vskip .13 in
\noindent{\bf \S~3. Mixed State Case}

In this section an individual system is a mixed state $\rho$ :

\eqn\nineteen{
\eqalign{
\rho &= \sum_i \left| \psi_i \right\rangle \pi_i \left\langle
\psi_i \right| \ ,\cr
\sum_i \pi_i &= 1 \ , \qquad \left\langle \psi_i |
\psi_i \right\rangle = 1\ .\cr}
}
The total system is written by the density matrix,

\eqn\twenty{
\rho^N = \sum_{i^1 \cdots i^N} \left| \psi_{i^1} \right\rangle
\cdots \left| \psi_{i^N} \right\rangle  \pi_{i^1} \cdots \pi_{i^N}
\left\langle \psi_{i^N} \right| \cdots \left\langle \psi_{i^1}
\right| \ .
}
Here upper indices of $i$ distinguish the individual system.
In the following discussion we assume the decoherence :

\eqn\twentyone{
\eqalign{
D\left(C_{\alpha^\prime}, C_\alpha\right) &=
\sum_i \pi_i \left\langle \psi_i \left| C^\dagger_\alpha
C_{\alpha^\prime} \right| \psi_i \right\rangle \cr
&= \delta_{\alpha\alpha^\prime} \sum_i \pi_i
\left\langle \psi_i \left| C^\dagger_\alpha C_\alpha \right|
\psi_i \right\rangle \ ,\cr}
}
where we have used Eqs.~\four, \six, \nineteen. We find

\eqn\twentytwo{
\eqalign{
P\left[C_\alpha\right] &= \sum_i \pi_i
\left\langle \psi_i \left| C^\dagger_\alpha C_\alpha
\right| \psi_i \right\rangle \cr
&= \sum_i \pi_i \left\langle \psi_i \left| C_\alpha \right|
\psi_i \right\rangle
= \sum_i \pi_i \left\langle \psi_i \left| C^\dagger_\alpha \right|
\psi_i \right\rangle \ , \cr}
}
using Eqs.~\two, \three, \eight, \twentyone. The Eqs.~\ten \ hold in
this case, too. We will show that $P\left[C_\alpha \right]$ are
equal to the relative frequencies.

Defining

\eqn\twentythree{
\left| \alpha , i \right\rangle = {C_\alpha \left| \psi_i \right\rangle
\over \sqrt{\sum_j \pi_j \left\langle \psi_j \left| C_\alpha
\right| \psi_j \right\rangle}}
\ ,
}
we obtain from Eqs.~\two, \three, \twentyone, \twentytwo \ that

\eqn\twentyfour{
\eqalign{
\sum_i \pi_i \left\langle \alpha , i | \beta , i
\right\rangle &= \delta_{\alpha \beta}\ , \cr
\left| \psi_i \right\rangle &= \sum_\alpha \sum_j
\pi_j \left\langle \alpha , j | \psi_j \right\rangle
\left| \alpha , i \right\rangle \ , \cr
\left| \sum_i \pi_i \left\langle \alpha , i | \psi_i
\right\rangle \right|^2 &= P \left[ C_\alpha \right] \ . \cr}
}
We can expand $\rho^N$ by these vectors $\left| \alpha , i
\right\rangle$ and write

\eqn\twentyfive{
\eqalign{
\rho^N &= \sum_{i^1 \cdots i^N} \left| \Psi_{i^1 \cdots i^N}
\right\rangle \pi_{i^1} \cdots \pi_{i^N}
\left\langle \Psi_{i^1 \cdots i^N} \right| \ , \cr
\left| \Psi_{i^1 \cdots i^N} \right\rangle &=
\left| \psi_{i^1} \right\rangle \cdots \left| \psi_{i^N}
\right\rangle \cr
&= \sum_{\scriptstyle \alpha^1 \cdots \alpha^N \atop
\scriptstyle j^1 \cdots j^N}
\pi_{j^1} \cdots \pi_{j^N}
\left\langle \alpha^1 , j^1 | \psi_{j^1} \right\rangle
\cdots \left\langle \alpha^N , j^N | \psi_{j^N} \right\rangle\cr
&\qquad \qquad \qquad \times
\left| \alpha^1 , i^1 \right\rangle \cdots
\left| \alpha^N , i^N \right\rangle \ . \cr}
}

Let us define

\eqn\twentysix{
\eqalign{
\left| \Psi^\epsilon_{i^1 \cdots i^N} \right\rangle &=
\sum_{{\scriptstyle \alpha^1 \cdots \alpha^N \atop
\scriptstyle \delta ( \alpha^1 \cdots \alpha^N )
< \epsilon}
\atop \scriptstyle j^1 \cdots j^N}
\pi_{j^1} \cdots \pi_{j^N}
\left\langle \alpha^1 , j^1 | \psi_{j^1} \right\rangle \cdots
\left\langle \alpha^N , j^N | \psi_{j^N} \right\rangle \cr
&\qquad \qquad \qquad \qquad \times
\left| \alpha^1 , i^1 \right\rangle \cdots
\left| \alpha^N , i^N \right\rangle \ , \cr
\left| \chi^\epsilon_{i^1 \cdots i^N} \right\rangle &=
\sum_{{\scriptstyle \alpha^1 \cdots \alpha^N \atop
\scriptstyle \delta ( \alpha^1 \cdots \alpha^N )
\ge \epsilon}
\atop \scriptstyle j^1 \cdots j^N}
\pi_{j^1} \cdots \pi_{j^N}
\left\langle \alpha^1 , j^1 | \psi_{j^1} \right\rangle \cdots
\left\langle \alpha^N , j^N | \psi_{j^N} \right\rangle \cr
&\qquad \qquad \qquad \qquad \times
\left| \alpha^1 , i^1 \right\rangle \cdots
\left| \alpha^N , i^N \right\rangle \ . \cr}
}
Again we can prove from Eqs.~\ten, \eleven, \twelve, \twentyfour,
\twentysix \ and Eq.~(4.13) of \Ref \refs{\DeWb}  that

\eqn\twentyseven{
\sum_{i^1 \cdots i^N} \pi_{i^1} \cdots \pi_{i^N}
\left\langle \chi^\epsilon_{i^1 \cdots i^N} |
\chi^\epsilon_{i^1 \cdots i^N} \right\rangle
\le {1 \over N\epsilon} \ .
}
Now if we assume

\eqn\twentyeight{
\left\langle \chi^\epsilon_{i^1 \cdots i^N} |
\chi^\epsilon_{i^1 \cdots i^N} \right\rangle \ge c \qquad
\left( \exists c > 0 \right) \ ,
}
when $N \to \infty$ , then Eq. \twentyseven \ means

\eqn\twentynine{
\sum_{i^1 \cdots i^N} \pi_{i^1} \cdots \pi_{i^N} c
\le {1 \over N\epsilon} \qquad
( N \to \infty ) \ .
}
{}From Eqs. \nineteen, \twentynine \ we obtain that

\eqn\thirty{
c \le {1 \over N\epsilon} \qquad
( N \to \infty ) \ ,
}
but this is a contradiction to $c > 0$ . Hence

\eqn\thirtyone{
\lim_{N \to \infty}
\left\langle \chi^\epsilon_{i^1 \cdots i^N} |
\chi^\epsilon_{i^1 \cdots i^N} \right\rangle = 0 \ .
}
This Eq.~\thirtyone \ means that in the expansion \twentyfive \
 we only need such $\alpha^I$ that satisfy
$\delta \left( \alpha^1, \cdots ,\alpha^N \right) < \epsilon$ ,
if we consider the limit of $N \to \infty$ .
So we have confirmed that $P\left[ C_\alpha \right]$ are
equal to the relative frequencies.

\vskip .13 in
\noindent{\bf \S~4. Summary}

In order to confirm the physical probability interpretation
of the Gell-Mann and Hartle's generalized quantum theory,
we started from $N$ identical systems, which were pure
states (\S 2) or mixed states (\S 3). We found that the
relative frequencies of histories $C_\alpha$ are equal to
the diagonal elements of decoherence functional
$P\left[C_\alpha \right]$, when the set of alternative histories
decoheres and $N \to \infty$.

\vskip .13 in
\noindent{\bf Acknowledgements}

The author would like to thank Professor J.B. Hartle for valuable
discussions and careful reading of the manuscript. He is also
grateful to Professor G.T. Horowitz for helpful discussions.
He would like to thank University of California, Santa Barbara for
hospitality. This work was supported in part by Japanese Ministry
of Education, Science and Culture.

\listrefs

\end